\documentclass[letterpaper, 10 pt, conference]{ieeeconf}  

\IEEEoverridecommandlockouts                              

\usepackage{amsmath}
\usepackage{amssymb}
\newtheorem{definition}{Definition}[section]
\newtheorem{proposition}{Proposition}[section]
\newtheorem{example}{Example}[section]
\newtheorem{problem}{Problem}[section]
\newtheorem{lemma}{Lemma}[section]

\title{\LARGE \bf
Bounding the $l_2$ sensitivity for positive linear observers*
}

\author{Aisling McGlinchey$^{1}$ and Oliver Mason$^{2}$
\thanks{Presented at the European Control Conference, 2018}
\thanks{*This work was supported with the financial support of the Science Foundation Ireland grant 13/RC/2094 and co-funded under the European Regional Development Fund through the Southern \& Eastern Regional Operational Programme to Lero - the Irish Software Research Centre (www.lero.ie)}
\thanks{$^{1}$Dept. of Mathematics and Statistics, Maynooth University, Maynooth, Co. Kildare, Ireland \& Lero, the Irish Software Research Centre {\tt\small aisling.mcglinchey.2009@mumail.ie}}%
\thanks{$^{2}$Dept. of Mathematics and Statistics/Hamilton Institute, Maynooth University, Maynooth, Co. Kildare, Ireland \& Lero, the Irish Software Research Centre {\tt\small oliver.mason@mu.ie}}%
}

\begin{document}
\maketitle

\thispagestyle{empty}
\pagestyle{empty}

\begin{abstract}
We consider the design of differentially private Luenberger observers for positive linear systems.  In particular, we derive a bound for the $l_2$ sensitivity of Luenberger observers, which is used to quantify the noise required to achieve relaxed differential privacy via the Gaussian mechanism.  An approach to minimise this bound for positive observers is described and several bounds relevant to this problem are derived. 
\end{abstract}

\section{Introduction}
Modern applications of control theory such as intelligent transportation and smart buildings rely on signals that contain personal information on users of the system.  For instance, in both of the above examples, it is not hard to see how location data plays a key role.  Concerns over data privacy in the general public and among regulatory authorities have thus necessitated the incorporation of formal privacy guarantees into the design of control systems.  Arguably the leading mathematical framework for privacy-preserving control systems is that provided by \emph{Differential Privacy} \cite{Dwork06}.  

Notwithstanding the importance of the issue, it is only in the recent past that researchers have begun to seriously address privacy as a design consideration in control theory; see \cite{LeNy3} for a recent overview of work on this topic.  In the recent past, the development of privacy-preserving solutions of various control problems have been considered.  In particular, the design of privacy preserving routing algorithms has been addressed in \cite{Don15}, while the design of differentially private algorithms for convex optimisation was addressed in \cite{Han} (and elsewhere).  The use of Gaussian mechanisms for allowing multiple data owners to make data available to multiple users in a differentially private way was studied in \cite{MoMuP}.  The central issue of the utility-privacy trade-off in a system theoretic setting was considered in the paper \cite{Ent} with utility being quantified using the information theoretic concept of entropy.   

The line of work initiated in \cite{LeNy1, LeNy2} inspires and motivates our discussion here.  In particular, in \cite{LeNy1}, the Laplace and Gaussian mechanisms were extended to a system-theoretic setting and results were given establishing how these could be used to design differentially private mechanisms for various control questions.  Specifically, Kalman filtering was considered in \cite{LeNy1} while the design of differentially private Luenberger observers was studied in \cite{LeNy2}.  As for static databases, the concept of \emph{sensitivity} is critical in determining the quantity of noise required to achieve a desired level of differential privacy.  

Our work in this paper focuses on differentially private observer design for the class of \emph{positive systems} \cite{FarRin, ForVal12,BlaColVal15}.  Such systems have attracted much attention for some time now and are relevant to applications in areas such as population dynamics, transportation and public health: privacy is an important consideration for all of these applications.  It is now well recognised that the design of positive linear observers for positive linear systems has a distinct character to the question for general systems \cite{ARami06}, \cite{BackAst08,Shu08}, \cite{VSch07}.  Similarly, when considering differentially private positive observer design, several novel questions, specific to the class of positive systems, naturally arise.  In a recent paper \cite{AMcG2017}, the authors considered the $l_1$ sensitivity of a positive Luenberger observer and derived bounds for this as well as presenting a simple algorithm for minimising these bounds where a single output is used to construct the observer.  In the current work, we will extend this analysis to the $l_2$ sensitivity, which is key for determining what level of noise is required for achieving relaxed differential privacy using the Gaussian mechanism.     

\subsection{Outline of paper}

In Section \ref{Sec:DP}, we introduce our notation and provide background on differential privacy for a dynamical system.  In Section \ref{Sec:sens}, we derive an upper bound for the $l_2$ sensitivity of a linear observer of Luenberger type.  In Section \ref{Sec:optObs}, we focus on positive systems and describe the steps of an algorithm for designing a positive observer that minimises the bound derived in Section \ref{Sec:sens}.

\section{Differential Privacy and Systems Theory}
\label{Sec:DP}
Throughout the paper, $\mathbb{R}^{n}$ denotes the vector space of $n$-tuples of real numbers and $\mathbb{R}^{m \times n}$ the space of $m \times n$ matrices with real entries.  For $x \in \mathbb{R}^n$: $x \geq 0$ means that $x_i \geq 0$ for $1 \leq i \leq n$.  $\mathbb{R}^n_+$ denotes the nonnegative orthant 
$$\mathbb{R}^n_+ := \{x \in \mathbb{R}^n \mid x \geq 0\}.$$
Analogous notation is also used for matrices throughout.  Also the notation $A \geq B$ denotes that $A-B \geq 0$ is a nonnegative matrix. $A^T$ denotes the transpose of $A \in \mathbb{R}^{n \times n}$ and $\rho(A)$ is used to denote its spectral radius.

We will work exclusively with the $l_2$ norm and the corresponding induced norm for matrices, which is commonly referred to as the \emph{spectral norm}.  We consider these norms as they are the natural norms for use in connection with Gaussian mechanisms for differential privacy \cite{LeNy1,Dwork06}.  As no confusion will arise, we use $\|x\|$ for the $l_2$ norm of $x \in \mathbb{R}^n$ while for $M \in \mathbb{R}^{m \times n}$, $\|M\|$ denotes the $l_2$ induced (spectral) norm of $M$.  It is standard \cite{HJ1} that this is given by 
\begin{equation}
\label{eq:l2colsum} \|M\| = \sqrt{\rho(M^TM)}.
\end{equation}

The following monotonicity property of the spectral norm will prove useful later on.
\begin{lemma}\label{lem:monspec}
Let $A, B$ in $\mathbb{R}^{m \times n}_+$ be given with $A \leq B$.  Then $\|A\| \leq \|B\|$.
\end{lemma}

We will consider general, discrete-time systems given by mappings between spaces of real-valued sequences.  Formally, a system $G:\mathcal{S}_1 \rightarrow \mathcal{S}_2$ maps from a space $\mathcal{S}_1$ consisting of sequences, each term of which lies in $\mathbb{R}^n$ for some $n \in \mathbb{N}$, to $\mathcal{S}_2$ consisting of sequences whose terms lie in $\mathbb{R}^p$ for some $p \in \mathbb{N}$.  

We assume that there is an adjacency relation (which is reflexive and symmetric), $u \sim u'$, defined on $\mathcal{S}_1$. The precise definition of $\sim$ depends on the context and reflects those changes in the input sequence that privacy preserving mechanisms should render difficult to detect.  

Given $\epsilon > 0$, $\delta > 0$, and a system $G:\mathcal{S}_1 \rightarrow \mathcal{S}_2$, an $(\epsilon, \delta)$-differentially private mechanism is defined by specifying a set of random variables $\{Z_{G, u} \mid u \in \mathcal{S}_1\}$ taking values in $\mathcal{S}_2$ satisfying:
\begin{equation}
\label{eq:DPSys}
\mathbb{P}(Z_{G, u} \in A) \leq e^{\epsilon} \mathbb{P}(Z_{G, u'} \in A) + \delta
\end{equation}
for all measurable sets $A$ of $\mathcal{S}_2$ and all $u \sim u'$. 

One way to achieve  $(\epsilon, \delta)$-differentially private mechanism is by adding \textit{i.i.d} Gaussian noise to each component of the output \cite{LeNy1}.  Following from the results from \cite{LeNy1, LeNy2} first, we need the $\mathcal{Q}$-function defined as $\mathcal{Q}(x) := \frac{1}{\sqrt{2 \pi}} \int_x^{\infty} \exp(-\frac{u^2}{2}) du $. Now for $\epsilon>0$, $0 \leq \delta \leq 0.5$, let $K = \mathcal{Q}^{-1}(\delta)$ and define $\kappa_{\delta,\epsilon} = \frac{1}{2\epsilon}(K+\sqrt{K^2+2\epsilon})$. Let the system $G$ with $m$ inputs and $p$ outputs be given.  The mechanism $M(u) = Gu +w$, where $w_k$ is white Gaussian noise with covariance matrix $\kappa^2_{\delta,\epsilon} ( \Delta G)^2 I_p$, is $(\epsilon, \delta)$-differentially private. The quantity $\Delta G$ appearing here is the $l_2$ sensitivity which we will define shortly. 


The $l_2$ norm of a sequence $y = (y(k))$, $0 \leq k < \infty$ with $y(k) \in \mathbb{R}^n$ for all $k$ is given by 
\begin{eqnarray}
\label{eq:norm}
\|y\|^2 = \sum_{k=0}^{\infty}\|y(k)\|^2
\end{eqnarray}

Now we recall the key concept of sensitivity which determines the magnitude of noise required to achieve differential privacy using the Gaussian mechanism. 
\begin{definition}
The $l_2$ sensitivity $\Delta(G)$ of a system $G$ is defined as 
\begin{equation}
\label{eq:sen1} \Delta(G) := \sup_{y \sim y'} \|G(y) - G(y')\|.
\end{equation}
\end{definition}
It is important to note that $\Delta(G)$ depends on both the norm and the adjacency relation.  For the remainder of this paper, our focus is on deriving bounds for the $l_2$ sensitivity of observers for linear systems, with particular focus on positive linear systems. 

\section{Sensitivity for Luenberger observers}
\label{Sec:sens}
In this section, we consider Luenberger observers for linear systems and with a view to the design of $(\epsilon, \delta)$ differentially private observers, we shall derive a bound on the $l_2$ sensitivity of such observers.  

Consider a system of the form 
\begin{eqnarray}
\label{eq:sys}
x(k+1) = Ax(k) \nonumber \\
y(k) = Cx(k) 
\end{eqnarray}
where $A \in \mathbb{R}^{n \times n}$, $C \in \mathbb{R}^{p \times n}$.  We will assume $p \leq n$ throughout.  Recall that to design an observer of Luenberger form requires us to construct a matrix $L \in \mathbb{R}^{n \times p}$ such that the solution $z(\cdot )$ of the system $\mathcal{L}$ given by 
\begin{eqnarray}
\label{eq:obs}
z(k+1) &=& Az(k) + L(y(k)-Cz(k)) \\
&=& (A-LC)z(k) +Ly(k) \nonumber 
\end{eqnarray}
satisfies $\| z(k) - x(k) \| \to 0 $ as $k \to \infty$ where $x$ is the solution to \eqref{eq:sys}. 

The output $y$ may contain sensitive or personal information. In this circumstance, we release a noisy observer $\hat{z}$ that is $(\epsilon, \delta)$ differentially private. 

For this paper, we will work with the following definition of adjacency from \cite{LeNy2}.

Let $K>0$ and $0 \leq \alpha < 1$ be given. Then two sequences $y, y'$ are adjacent, $y \sim y'$ if 
\begin{eqnarray}
\label{eq:adj} \exists \, k_0 \geq 0 \, s.t. 
	\begin{cases} 
		y(k)= y'(k),&k<k_0\\ 
		\|y(k)-y'(k)\|\leq K\alpha^{k-k_0},& k\geq k_0 
	\end{cases}\\ \nonumber
\end{eqnarray}
This adjacency definition corresponds to a change in signals due to a small number of people at time $k_0$. The initial magnitude of the change is $K$ and it decays geometrically at a rate of $\alpha$. 

\subsection{Bounding the sensitivity}
In this subsection, we first derive an upper bound of the $l_2$ sensitivity of the system \eqref{eq:obs}; later we shall consider the problem of minimising this bound for positive observers. 
\begin{proposition}
\label{prop:SenBd} Let $A \in \mathbb{R}^{n \times n}$, $C \in \mathbb{R}^{p \times n}$, $L \in \mathbb{R}^{n \times p}$ be given and suppose that $\|A-LC\| < 1$.  Consider the Luenberger observer, $\mathcal{L}$ given by \eqref{eq:obs}.  Then for the adjacency relation defined by \eqref{eq:adj}, the $l_2$ sensitivity of $\mathcal{L}$ (as given in (\ref{eq:sen1})) is bounded by
\begin{equation}\label{eq:senbd1} 
\Delta(\mathcal{L})^2 \leq \frac{K^2}{1-\alpha^2} \left( \frac{1+N\alpha}{1-N\alpha} \right) \left( \frac{\|L\|^2}{1-N^2} \right)
\end{equation}
where $N = \|A-LC\|$.
\end{proposition}

\textbf{Proof:} Let two adjacent sequences $y \sim y'$ be given. The (zero-initial state) map from $y$ to $z$ and $y'$ to $z'$ corresponding to \eqref{eq:obs} is described by
\begin{eqnarray}
\label{eq:obs2} z(k+1) = \sum_{i=0}^{k} (A-LC)^{k-i}L y(i) 
\end{eqnarray}
\begin{eqnarray}
\label{eq:obstilde} z'(k+1) = \sum_{i=0}^{k} (A-LC)^{k-i}Ly'(i). 
\end{eqnarray}
As $y \sim y'$, it follows that $z'(k)= z(k)$ for $k\leq k_0$. Moreover, for $k> k_0$, we have 
\begin{eqnarray}
z(k)-z'(k) = \sum_{i=k_0}^{k-1} (A-LC)^{k-i-1} L(y(i)-y'(i)).
\end{eqnarray}
We need to bound $\|z(k)-z'(k)\|$; for $k > k_0$.  Using the triangle inequality and the submultiplicative property of the spectral norm we have:
\begin{eqnarray*}\nonumber
\|z(k)- z'(k)\| &=& \| \sum_{i=k_0}^{k-1} (A-LC)^{k-i-1} L(y(i)-y'(i)) \| \\
&\leq&  \|L\| K \sum_{i=k_0}^{k-1} \|A-LC\|^{k-1-i} \alpha^{i-k_0}
\end{eqnarray*}
To simplify the calculations let $\|A-LC\| = N$.  The $l_2$ norm of $z - z'$ is given by \eqref{eq:norm} which applied to the above calculation shows that $\|z-z'\|^2$ is bounded above by:
\begin{eqnarray}
\|L\|^2 K^2 \sum_{k=k_0+1}^{\infty} \left(\sum_{i=k_0}^{k-1} N^{k-1-i} \alpha^{i-k_0}\right)^2 \nonumber
\end{eqnarray}
A simple shift of indices shows that the series in the above expression is equal to: 
\begin{eqnarray}\label{eq:Series1}
\sum_{k=0}^{\infty} \left(\sum_{i=0}^{k} N^{k-i} \alpha^{i}\right)^2 
\end{eqnarray}
In order to evaluate the above sum, we shall show that the series  $\sum_{k=0}^{\infty} \left(\sum_{i=0}^{k} N^{k-i} \alpha^{i}\right)^2$ is equal to the following sum of two series: 
\begin{eqnarray} \label{eq:Series2}
\sum_{i=1}^{\infty} i\alpha^{i-1} \frac{N^{i-1}}{1-N^{2}} + \sum_{i=1}^{\infty}i N^{i-1}\frac{\alpha^{i+1}}{1-\alpha^2} 
\end{eqnarray}
To make the argument a little easier to follow, we note the pattern of the terms $\left(\sum_{i=0}^{k} N^{k-i} \alpha^{i}\right)^2$ for $k=0,1,2,3,4$. 
\begin{gather*}
1  \nonumber \\
\label{eq:Ser2} N^2 + 2N\alpha +\alpha^2 \\
N^4 + 2N^3\alpha + 3N^2\alpha^2 + 2N\alpha^3 + \alpha^4 \nonumber \\
N^6 + 2N^5\alpha + 3N^4\alpha^2 + 4N^3\alpha^3 + 3N^2\alpha^4 + 2N\alpha^5 + \alpha^6 \nonumber \\
N^8 + 2N^7\alpha + 3N^6\alpha^2 + 4N^5\alpha^3 + 5N^4\alpha^4 + 4N^3\alpha^5  \nonumber \\
+ 3N^2\alpha^6 + 2N\alpha^7 + \alpha^8 \nonumber
\end{gather*}
To see the above identity, we need to make the following observations.
\begin{itemize}
\item The total power of each monomial term $N^p \alpha^q$ in $\left(\sum_{i=0}^{k} N^{k-i} \alpha^{i}\right)^2$ is $2k$.  This implies that each monomial in \eqref{eq:Series1} appears for exactly one value of $k$.
\item A straightforward calculation shows that each term in \eqref{eq:Series1} is either of the form $N^p \alpha^{p+2s}$ for some integers $p \geq 0$, $s > 0$ or of the form $\alpha^p N^{p + 2s}$ for integers $p \geq 0$, $s \geq 0$.  
\item Finally, it is not too difficult to verify that the coefficient of the monomial $N^p \alpha^q$ in $\left(\sum_{i=0}^{k} N^{k-i} \alpha^{i}\right)^2$ is given by $\min\{p, q\}+1$. 
\end{itemize}
From the above observations, it follows that we can split the terms in \eqref{eq:Series1} into those of the form $i(\alpha^{i-1} N^{(i-1)+2s})$ for $1 \leq i < \infty$, $s \geq 0$ and those of the form $i(N^{i-1} \alpha^{(i-1)+2s})$ for $1 \leq i < \infty$, $s > 0$.  The equality of \eqref{eq:Series1} and \eqref{eq:Series2} now follows from the above points.  

Thus, the series \eqref{eq:Series1} can now be rearranged as follows:
\begin{eqnarray*}
\sum_{i=1}^{\infty} i\alpha^{i-1} \frac{N^{i-1}}{1-N^{2}} + \sum_{i=1}^{\infty}i N^{i-1}\frac{\alpha^{i+1}}{1-\alpha^2} \\
= \frac{1}{1-N^2} \sum_{i=1}^{\infty} i\alpha^{i-1} N^{i-1} + \frac{\alpha^2}{1-\alpha^2} \sum_{i=1}^{\infty} iN^{i-1}\alpha^{i-1}  \\
=  \left( \frac{1}{1-N^2} + \frac{\alpha^2}{1-\alpha^2} \right) \sum_{i=1}^{\infty} i(N\alpha)^{i-1}  \\
= \left( \frac{1}{1-N^2} + \frac{\alpha^2}{1-\alpha^2} \right) \sum_{i=1}^{\infty} i\beta ^{i-1}  
\end{eqnarray*}
where $\beta = N\alpha < 1$.
To evaluate $\sum_{i=1}^{\infty} i\beta^{i-1}$, we simply note that it is the derivative of the absolutely convergent (for $|\beta| < 1$) power series $\sum_{i=0}^{\infty} \beta^i$ and thus
\begin{eqnarray*}
\sum_{i=1}^{\infty} i\beta^{i-1} &=& \frac{d}{d\beta} \left( \sum_{i=0}^{\infty} \beta^i \right)  \\
&=& \frac{d}{d\beta} \left(\frac{1}{1-\beta} \right)  \\
&=& \frac{1}{(1-\beta)^2} \\ 
&=& \frac{1}{(1-N\alpha)^2}.
\end{eqnarray*}
Therefore, we may write
\begin{eqnarray*}
\sum_{k=0}^{\infty} \left(\sum_{i=0}^{k} N^{k-i} \alpha^{i}\right)^2 &=& \left( \frac{1}{1-N^2} + \frac{\alpha^2}{1-\alpha^2} \right) \sum_{i=1}^{\infty} i\beta^{i-1} \\
&=& \left( \frac{1}{1-N^2} + \frac{\alpha^2}{1-\alpha^2} \right) \frac{1}{(1-N\alpha)^2}  \\
&=& \frac{1}{(1-N\alpha)^2} \left( \frac{1-N^2\alpha^2}{(1-N^2)(1-\alpha^2)} \right) \\
&=& \frac{1}{1-\alpha^2} \left( \frac{1+N\alpha}{1-N\alpha} \right) \left( \frac{1}{1-N^2} \right)
\end{eqnarray*}
Combining everything, the square of the $l_2$ norm of $z - z'$ is bounded above by:
\begin{eqnarray*}
\frac{K^2}{1-\alpha^2} \left( \frac{1+N\alpha}{1-N\alpha} \right) \left( \frac{\|L\|^2}{1-N^2} \right)
\end{eqnarray*}
which completes the proof as $y\sim y'$ were arbitrary.

\vspace{5mm}

In the paper \cite{AMcG2017} the authors derived an upper bound for the $l_1$ sensitivity of a Luenberger observer. The bound in this case was given by:
\begin{equation}\label{eq:senbdl1} 
\Delta(\mathcal{L}) \leq \left(\frac{K}{1-\alpha}\right)\left(\frac{\|L\|_1}{1-\|A-LC\|_1}\right),
\end{equation} 
where the induced matrix norms are with respect to the $l_1$ vector norm.  This bound has a simpler form than our $l_2$ bound as we are able to isolate $K$ and $\alpha$; thus when seeking to minimise the $l_1$ bound, we only have to consider the function $$\frac{\|L\|_1}{1-\|A-LC\|_1}.$$

\subsection{Tightness of upper bound}
Here we describe a simple example to illustrate how the bound in Proposition \ref{prop:SenBd} can be attained for certain values of $A, L, C$.  It is worth noting that the matrices in the example are all nonnegative.   
\begin{example}
Consider 
$$A = \left(\begin{array}{c c}
		1/4 & 1/2\\
        1/2 & 1
		\end{array}\right), \,\,
C = \left(\begin{array}{c c}
			1/3 & 2/3
            \end{array}\right), \,\,
L = \left(\begin{array}{c}
			1/3\\
            2/3
            \end{array}\right).$$
A straightforward calculation shows that            
$$A-LC = \left(\begin{array}{c c}
			5/36 & 5/18\\
            5/18 & 5/9
            \end{array}
            \right).$$

Direct calculation shows that $\|A-LC\|=25/36<1$. Moreover, we can verify that $$(A-LC)L = (25/36)L = \|A-LC\|^2 L$$ so that $(A-LC)^i L = (\|A-LC\|^2) ^i L$ for all $i \geq 1$. Using this, we can see that for a pair of adjacent sequences $y, y'$ where $y(k), y'(k)$ are in $\mathbb{R}$ for all $k$ and satisfy the adjacency relation $y \sim y'$ with equality replacing the inequality (as each $y(k)$ is simply a real number, this is not difficult to achieve), the corresponding observer outputs $z, z'$ satisfy $$\|z-z'\| = \frac{K^2}{1-\alpha^2} \left( \frac{1+N\alpha}{1-N\alpha} \right) \left( \frac{\|L\|^2}{1-N^2} \right)$$

\end{example}

\section{Observer design and sensitivity bounds}
\label{Sec:optObs}
In applications such as transportation and public health the underlying dynamical system is typically a positive system.  When designing an observer for a positive system it is natural to require that all signals appearing in the observer system \eqref{eq:obs} are nonnegative \cite{ARami06,VSch07,BackAst08}.  The problem of positive observer design has attracted a considerable amount of attention and a variety of more general observer types than the simple class studied here have been considered \cite{Shu08,BackAst08}. 

As in \cite{AMcG2017}, our focus for the remainder of the paper is on a number of questions specific to the design of differentially private positive observers.  In particular, our discussion centres on the following fundamental question.
\begin{quote}
Construct a positive Luenberger observer with matrix $L$ that minimises the bound on the $l_2$ sensitivity given in Proposition \ref{prop:SenBd}. 
\end{quote}

The system \eqref{eq:sys} is positive if $A$ and $C$ are nonnegative. It has been shown in \cite{ARami06} that the existence of a positive observer is equivalent to the existence of a matrix $L$ such that $A-LC \geq 0$, $LC \geq 0$ and $\rho(A-LC) < 1$. As our interest is in the bound on the $l_2$ sensitivity of such an observer, we shall replace the requirement that $\rho(A-LC) < 1$ with $\|A-LC\| < 1$.  

\subsection{Minimising the sensitivity bound}
We first turn our attention to the question of characterising the infimal or minimal value of the bound in Proposition \ref{prop:SenBd} and of determining an observer matrix $L$ that either attains or approximates this value.  

As $K$ and $\alpha$ are fixed values determined by the adjacency \eqref{eq:adj} we wish to solve the following problem: 

\begin{problem}
\label{prob:L2Min}
Given $A \in \mathbb{R}_+^{n \times n}$, $C \in \mathbb{R}_+^{p \times n}$, consider
\begin{eqnarray}\label{eq:FL}
F(L) = \|L\|^2 \left( \frac{1+\|A-LC\| \alpha}{1-\|A-LC\| \alpha} \right) \left( \frac{1}{1- \|A-LC\|^2} \right).
\end{eqnarray}
Determine $\textrm{inf} \,F(L)$ subject to the constraints:
\begin{eqnarray}
\label{eq:feas1} LC \geq 0 ,\, A-LC \geq 0, \, \|A-LC\| < 1 
\end{eqnarray}
\end{problem}

From now on, we assume throughout that the set defined by \eqref{eq:feas1} is non-empty: in the case where it is empty, most statements are satisfied vacuously (or by adopting the convention that the infimum of an empty set is $\infty$).  In the remainder of this section, we shall present a number of results that provide a foundation for methods to minimise $F(L)$ subject to \eqref{eq:feas1}.  In particular, we shall show how to reduce this question to a simple 1-dimensional question via a family of optimisation problems with a simpler convex objective function.  Our results also provide insight into the fundamental trade-off between utility and privacy by clarifying the relationship between the achievable sensitivity values and observer convergence rates as specified by $\|A-LC\|$.   

In the next lemma, we note that in the case where $\|A\| > 1$, the infimum above is in fact a minimum that is attained for some $L$ in the feasible set.  

\begin{lemma}\label{lem:Min} Let $A \in \mathbb{R}^{n\times n}_+$, $C \in \mathbb{R}^{p\times n}_+$ be given with $\|A\| > 1$.  Then there exists some $\kappa$ with $0 < \kappa < 1$ such that $\textrm{inf} \, F(L)$ subject to \eqref{eq:feas1} is given by $\textrm{min} \, F(L)$ subject to 
\begin{eqnarray}
\label{eq:feas2} LC \geq 0 ,\, A-LC \geq 0, \, \|A-LC\| \leq \kappa
\end{eqnarray}
\end{lemma}

\textbf{Proof:}  We first note that for any $\kappa$ with $0 < \kappa < 1$, $F$ is a well-defined continuous function on the compact set given by \eqref{eq:feas2} and hence attains its minimum on this set.  

As $\|A\| > 1$, it follows that there must exist some $\epsilon > 0$ such that for any $L$ such that $\|A - LC \| < 1$, we have $\|L\| \geq \epsilon$.  It follows readily that for any $L$ satisfying \eqref{eq:feas1}, we must have 
\[F(L) \geq \frac{\epsilon^2}{1- \|A-LC\|^2}.\]
Now choose some $L_1$ satisfying \eqref{eq:feas1} and suppose $F(L_1) = f_1$.  If $f_1 = 0$, then clearly $F$ attains its minimum at $L_1$ and the result follows.  Otherwise, set $\epsilon_1 = \min \{\epsilon, \frac{f_1}{2} \}$.  Then a simple calculation shows that for any $L$ satisfying \eqref{eq:feas1} with $\|A-LC\| > 1 - \frac{\epsilon_1}{f_1}$, we must have $F(L) > f_1$.  The result now follows with $\kappa = 1 - \frac{\epsilon_1}{f_1}$.  

For the remainder of the paper, we shall assume that $\|A\| > 1$.  The case where $\|A\| < 1$ is trivial as we can simply take $L = 0$; the boundary situation where $\|A\| = 1$ may be more subtle however and require further analysis.

For notational convenience we will write $F(L) = \|L\|^2 \,\, H(\|A-LC\|)$, where $H:[0, 1) \rightarrow \mathbb{R}$ is given by
\begin{eqnarray}
\label{eq:H}
H(N) =\left( \frac{1+N \alpha}{1-N \alpha} \right) \left( \frac{1}{1-N^2} \right)
\end{eqnarray}
The next lemma will play a key role in our later approach to solving Problem \ref{prob:L2Min}.
\begin{lemma}
The function $H$ defined by \eqref{eq:H} is an increasing function on the interval $[0,1)$.
\end{lemma}

\textbf{Proof:} The first derivative of $H$ is:
\begin{eqnarray*}
H'(N) &=& \left(\frac{1}{1-N^2}\right) \left(\frac{(1+N\alpha)\alpha}{(1-N\alpha)^2} + \frac{\alpha}{1-N\alpha} \right) \\
&+& \left(\frac{1+N\alpha }{1-N\alpha}\right) \left(\frac{2N}{(1-N^2)^2}\right) 
\end{eqnarray*}
As $\alpha$ is a fixed value and $0 \leq \alpha \leq 1 $ it follows that $H'(N) > 0$ for $N$ in the interval $[0,1)$.

We will now outline the key steps of an algorithm for solving problem \ref{prob:L2Min}.  
\begin{enumerate}
\item \emph{Determine upper and lower bounds, $\eta_{min}$, $\eta_{max}$ for $\|L\|$ for $L$ satisfying \eqref{eq:feas1}}.
\begin{itemize}
\item For any system, we can clearly take 0 as a lower bound.  However, we shall shortly describe how to obtain a less conservative lower bound than this.
\item For a single output system ($p = 1$), an upper bound can be calculated as follows.  $L \in \mathbb{R}_+^{n \times 1}$ so we can consider $L$ as a column vector $l \in \mathbb{R}^n$; similarly we can take $C$ to be given by a row vector $c^T$ for $c \in \mathbb{R}^n$.  Since $A-lc^T \geq 0 $, $l_i \leq \frac{a_{ij}}{c_j} \implies l_i \leq \min_j \frac{a_{ij}}{c_j}$, for $1 \leq i \leq n$.  This gives an upper bound on the $l_2$ norm of $L$: namely 
\[\|L\| \leq \sqrt{\sum_i \left( \min_j \frac{a_{ij}}{c_j} \right)^2} = \eta_{max}. \]  Later we shall describe how to calculate an upper bound for general multiple output systems. 
\end{itemize}
\item \emph{For $\eta \in [\eta_{min},\eta_{max}]$ solve the optimisation problem $M(\eta) = \min H(N(L))$ such that $\|L\| = \eta$.}
\begin{itemize}
\item As $H$ is an increasing function of $N$, this is equivalent to solving the simpler problem of determining for each $\eta \in [\eta_{min},\eta_{max}]$:
\begin{eqnarray}
\label{eq:MinLNorm}
N(\eta) = \min_{\|L\|=\eta} \|A-LC\|
\end{eqnarray}
subject to $L$ satisfying \eqref{eq:feas1}.
\item This problem may be solved for each $\eta$ using optimisation packages based on techniques such as sequential quadratic programming (SQP).  We shall show below that for the single output case, this step is in fact a convex problem.  
\end{itemize}
\item Finally solve $\min \eta^2 H(N(\eta))$ for $\eta \in [\eta_{min}, \eta_{max}]$. 
\begin{itemize}
\item This is a simple single variable optimisation problem so we have reduced the original multivariable, non-convex problem to optimising a single variable function in a closed interval.   
\end{itemize}
\end{enumerate}

\subsection{Bounding $\|L\|$}
We next turn our attention to the question of determining lower and upper bounds for $\|L\|$ where $L$ satisfies \eqref{eq:feas1} and the matrix $C$ has full rank $p$ ($p \leq n$).  

\begin{proposition}\label{prop:upperbd}
Let $A \in \mathbb{R}^{n \times n}_+$, $C \in \mathbb{R}^{p \times n}_+$ ($p \leq n$) be given and suppose that $L \in \mathbb{R}^{n\times p}$ satisfies \eqref{eq:feas1}.  Assume that $C$ has rank $p$. Then 
\begin{equation}\label{eq:Lbd}
\frac{\|A\| - 1}{\|C\|} \leq \|L\| \leq \|A\| \|C^{\dagger}\|
\end{equation}
where $C^{\dagger}$ is the pseudo-inverse of $C$.
\end{proposition}
\textbf{Proof:} As $p \leq n$ and we are assuming that $C$ has rank $p$, it follows easily from considering the singular value decomposition of $C$ that the pseudo-inverse $C^{\dagger}$ of $C$ exists satisfying $CC^{\dagger} = I_p$ where $I_p$ is the $p\times p$ identity matrix.  From this, the upper bound above follows readily as 
\begin{eqnarray*}
\|L\| &=& \|LCC^{\dagger}\| \\
&\leq& \|LC\|\|C^{\dagger}\| \\
&\leq& \|A\| \|C^{\dagger}\|
\end{eqnarray*}
where the last inequality follows as $0 \leq LC \leq A$ and the monotonicity of the spectral norm on nonnegative matrices.  

For the lower bound, note that
\begin{eqnarray*}
\|A\| - \|LC\| \leq \|A-LC\| < 1
\end{eqnarray*}
which implies that 
\[\|L\|\|C\| \geq \|LC\| > \|A\| - 1. \]
The lower bound now follows immediately.  

\subsection{The single output case $p=1$}
In this subsection, we note that when $p = 1$, the optimisation in step 2 of our algorithm for minimising $F(L)$ is equivalent to a convex problem.   
\begin{lemma}\label{lem:scaling}
Let $A \in \mathbb{R}^{n \times n}_+$, $c \in \mathbb{R}^{p}_+$ and $\eta > 0$ be given where $\eta < \eta_{\max}$.  Define:
\begin{eqnarray*}
m_1 &:=& \min \{\|A-lc^T\| \mid l \geq 0, A-lc^T \geq 0, \|l\| = \eta\};\\
m_2 &:=& \min \{\|A-lc^T\| \mid l \geq 0, A-lc^T \geq 0, \|l\| \leq \eta\}.
\end{eqnarray*}
Then $m_1 = m_2$. 
\end{lemma}
\textbf{Proof:} It suffices to show that for any $l$ such that $lc^T \geq 0$, $A-lc^T \geq 0$ and $\|l\| \leq \eta$, there exists some $l_1$ with $l_1c^T \geq 0$, $A-l_1c^T \geq 0$ and $\|l_1\| = \eta$ such that 
\[\|A-l_1c^T\| \leq \|A-lc^T\|.\]
To this end, let such an $l$ be given.  Take $\hat{l}$ to be the vector with $\hat{l}_i = \min_j\frac{a_{ij}}{c_j}$ so that $l \leq \hat{l}$ for any $l$ satisfying \eqref{eq:feas1} and $\|\hat{l}\| \geq \eta$.  It is not difficult to see that $\hat{l}$ lies in the convex feasible set $\mathcal{F}$ (assumed to be non-empty) determined by \eqref{eq:feas1}.    
For $\alpha \in [0,1]$ consider \[\hat{l}(\alpha) = \alpha \hat{l} + (1-\alpha) l.\]  As $\mathcal{F}$ is convex, $\hat{l}(\alpha)$ is in $\mathcal{F}$ for all $\alpha \in [0, 1]$.  Moreover, as $\|\hat{l}(0)\| = \|l\| < \eta$ and $\|\hat{l}(1)\| = \|\hat{l}\| \geq \eta$, it follows that for some $\alpha_1$ we must have $\|\hat{l}(\alpha_1)\| = \eta$.  The result follows setting $l_1 = \hat{l}(\alpha_1)$.   

\vspace{5mm}

The equality of $m_1$ and $m_2$ above shows determining the value of $m_1$ is a convex optimisation problem for the case $p = 1$.

Now we will describe a simple 2-dimensional example where we seek to minimise $F(L)$ given by \eqref{eq:FL} for a given $A, C$.
\begin{example}
Consider 
$$A = \left(\begin{array}{c c}
		1/4 & 1/2\\
        1/2 & 1
		\end{array}\right), \,\,
C = \left(\begin{array}{c c}
			1/3 & 2/3
            \end{array}\right)$$
With an algorithm as described above $\eta_{max}=1.6771$ and $\epsilon = 0.2$ and $K=0.5$ we get $L = [0.47692 \,\,\, 0.95385]^T$, so $\Delta(\mathcal{L})^2 = 1.2958$.
\end{example}





\section{Conclusions and future work}
We have derived an upper bound for the $l_2$ sensitivity of a Luenberger observer for a linear system.  We have then considered the problem of minimising this bound for positive systems where we also require that the observer be positive.  In particular we have described an algorithm that reduces this problem to a single variable optimisation and have given upper and lower bounds for the variable in this simple problem.  When the system has a single output, we show that the key step in the minimisation algorithm is a convex problem and have given simple bounds for this case also.  

In the design of differentially private observers there is a clear trade-off between the sensitivity as described by the function $F(L)$ and the norm of $\|A-LC\|$ which determines the rate of convergence of the observer to the value of the state.  In future work, it would be interesting to characterise the interplay between these two quantities and perhaps to characterise pareto solutions for what is essentially a multi-objective optimisation in this instance.  Two very natural design problems suggest themselves in this context.

\begin{itemize}
\item The construction of a positive observer with minimal sensitivity for a prescribed performance level, expressed in terms of the norm of $A-LC$.
\item The construction of a positive observer with optimal convergence  for a specified level of sensitivity.
\end{itemize}

Formally, the first of these questions amounts to determining the minimal value of $F(L)$ subject to $\|A-LC\| = \eta$, $A- LC \geq 0$, $LC \geq 0$.  It is trivial to see that once $\|A-LC\| = \eta$, then 
\[F(L) = \|L\|^2 \frac{1+\alpha \eta}{1 - \alpha \eta}\frac{1}{1 - \eta^2}.\]
Thus the question of determining the minimum sensitivity amounts to finding $L$ such that $\|L\|$ is minimal while $\|A-LC\| = \eta$.  

Formally, the second question amounts to determining the minimum value of $\|A-LC\|$ subject to $F(L) = \eta$, $A-LC \geq 0$, $LC \geq 0$.  A related direction for work is to characterise the geometry of the set of sensitivity/convergence pairs possible for a given pair of matrices $A, C$. 


\bibliographystyle{unsrt}
\bibliography{bibliography}

\end{document}